\documentclass[conference]{IEEEtran}
\IEEEoverridecommandlockouts

\usepackage{xurl}
\usepackage[hidelinks]{hyperref}

\usepackage{colortbl}
\usepackage{cite}
\usepackage{amsmath,amssymb,amsfonts}
\usepackage{algorithmic}
\usepackage{graphicx}

\usepackage{subcaption}
\usepackage{multirow}
\usepackage{booktabs}
\usepackage{color,soul}
\usepackage{stix}
\usepackage[table]{xcolor}

\usepackage{tcolorbox}
\usepackage{dblfloatfix}
\usepackage{booktabs}

\newtcolorbox{boxH}{
    colback = white!90!gray, 
    colframe = black, 
    boxrule = 0pt, 
    leftrule = 3pt
}

\newtcolorbox{boxH2}{
    colback = brown!10, 
    colframe = black, 
    boxrule = 0pt, 
    leftrule = 3pt
}

\def\BibTeX{{\rm B\kern-.05em{\sc i\kern-.025em b}\kern-.08em
    T\kern-.1667em\lower.7ex\hbox{E}\kern-.125emX}}
\begin{document}

\title{AI Policy, Disclosure, and Human in the Loop: How Are Contribution Guidelines Adapting to GenAI?}

% \title{Transparency, Accountability, and Quality: How Are Contribution Guidelines Adapting to GenAI?}

 \author{
 \IEEEauthorblockN{Andre Hora}
 \IEEEauthorblockA{
\textit{Department of Computer Science, UFMG}\\
 Belo Horizonte, Brazil \\
 andrehora@dcc.ufmg.br}
 \and
  \IEEEauthorblockN{Romain Robbes}
 \IEEEauthorblockA{
\textit{Univ. Bordeaux, CNRS, Bordeaux INP, LaBRI}\\
 Bordeaux, France \\
 romain.robbes@labri.fr}
 }

% \author{\IEEEauthorblockN{Anonymous}
% \IEEEauthorblockA{\textit{dept. name of organization} \\
% \textit{name of organization}\\
% City, Country \\
% email address}
% }

\maketitle

\begin{abstract}
Generative AI (GenAI) has recently transformed software development.
Due to the ease of generating code, open source projects are experiencing a growth in contributions.
To address the rise of GenAI, open source projects have begun implementing policies for AI usage in contributions.
However, the extent to which open source specifies whether AI-assisted contributions are allowed or prohibited, along with the best practices for contributors, remains unclear.
This paper provides an initial empirical study to explore how open source projects are adapting to GenAI contributions.
We analyzed 1,000 popular GitHub repositories and identified 118 AI policies for contributors.
Our results show that 
(1) 78\% of the AI policies allow AI-assisted contributions, while 22\% explicitly discourage AI use.
(2) 51\% of the AI policies require the disclosure of AI-assisted contributions; and
(3) 74\% of the AI policies require a human in the loop during contribution.
Overall, we find that the majority of the analyzed AI policies are positive regarding the usage of GenAI.
However, AI disclosure and human in the loop are fundamental in the contribution process.
Finally, we conclude by discussing implications for developers and researchers.
\end{abstract}

% Maybe for later, or is it the same thing: Policies vs process?
% for instance, the pi coding agent auto-closes PRs and issues:
% https://github.com/earendil-works/pi/pull/4409#issuecomment-4420884060
% this issue was closed, then re-opened:
% https://github.com/earendil-works/pi/issues/4390#issuecomment-4417500389

% some projects with extreme amounts of PRs, bugs, may have interesting policies. 
% e.g., OpenClaw:
% https://github.com/openclaw/openclaw/pull/80637
% and this one especially:
% https://github.com/openclaw/openclaw/issues/38283


\section{Introduction}

Generative AI (GenAI) has transformed how software is designed, developed, and engineered~\cite{fan2023large, hou2023large}.
Coding agents such as Claude Code, Cursor, and Codex have seen rapid recent adoption, operating with a high degree of \emph{autonomy}; for example, they can invoke external tools, execute code, and complete end-to-end development tasks~\cite{agentminingpaper}.
A recent large-scale study estimates that coding agent adoption on GitHub is close to 30\%~\cite{robbes2026agentic} and 76\% in new projects~\cite{robbes2026verymuchagentic}.

Due to the ease of generating code, open source projects are experiencing a growth in contributions~\cite{song2024impact, nakashima2026agentic, yang2026beyond, baltes2026endless}.
The volume contributions partially or entirely AI-generated have outpaced the maintenance capacity of many projects, creating debates within the open source community~\cite{yang2026beyond, llvm-1, llvm-2, rust-lang}.
As a result, maintainers often lack sufficient capacity to review submissions effectively~\cite{yang2026beyond}.
AI-assisted contributions may also lack quality—a phenomenon known as AI slop, which refers to low-quality content produced at scale using AI~\cite{ai-slop, baltes2026endless, baltes2026ai}.

\begin{figure}[t]
    \centering
    \fbox{\includegraphics[width=0.48\textwidth]{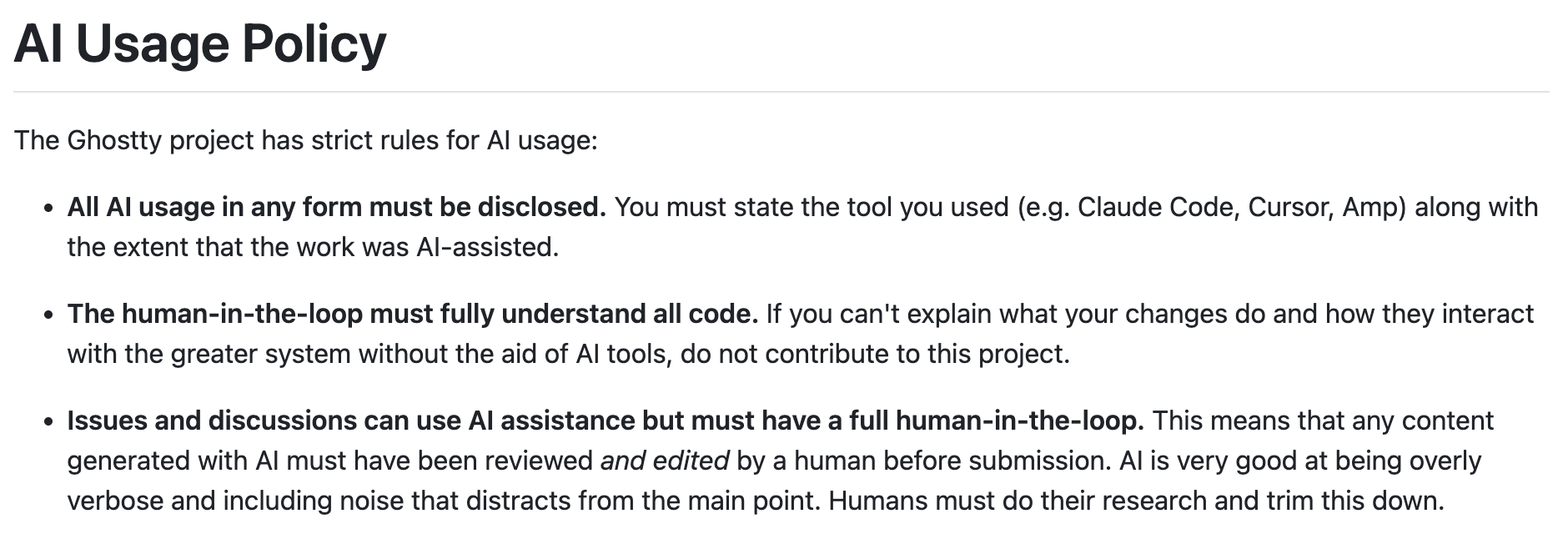}}
    \caption{AI policy of project ghostty-org/ghostty.}
    \label{fig:genai-ghostty}
\end{figure}

To address the rise of generative AI, open source projects have begun implementing policies for AI usage in contributions.
Several open source projects have established AI policies, including Linux~\cite{linux-policy}, Fedora~\cite{fedora-policy}, and LLVM~\cite{llvm-policy}.
Figure~\ref{fig:genai-ghostty} presents the AI policy of the Ghostty project, which states that AI-generated contributions are allowed provided they are disclosed and include human involvement in the contribution process~\cite{ghostty-org/ghostty}.
FastAPI contribution guidelines state that low-effort or low-quality pull requests are not accepted, and that AI-generated submissions may be closed and accounts blocked~\cite{fastapi-fastapi-ai-policy}.
Some projects are more welcoming toward generative AI.
For instance, the AI policy of huggingface/transformers states that ``\emph{AI-assisted contributions are welcome}''~\cite{huggingface-transformers}.
In contrast, other projects are more restrictive, for example, the 9001/copyparty policy states: ``\emph{Do not use AI / LLM when writing code}''~\cite{9001-copyparty}.

However, the extent to which open source projects specify whether AI-assisted contributions are allowed or prohibited, along with the best practices contributors are expected to follow, remains unclear.

% https://github.com/immich-app/immich/blob/a3ee615c5b2b3fb7cbc126547c96e8346b87bf59/CONTRIBUTING.md
% Use of generative AI; We ask you not to open PRs generated with an LLM. We find that code generated like this tends to need a large amount of back-and-forth, which is a very inefficient use of our time. 

This short paper provides an initial empirical study to explore how open source projects are adapting to generative AI contributions.
We analyzed 1,000 popular GitHub repositories and identified 118 AI policies for contributors.
We propose three research questions to address AI policy, AI disclosure, and human involvement in contributions:

\begin{itemize}
    \item \textbf{RQ1. How do open source projects address the use of generative AI in contributions?}
    Overall, 78\% of the AI policies allow AI-assisted contributions, while 22\% explicitly discourage AI use.
    
    \item \textbf{RQ2. How do open source projects address disclosure of AI-assisted contributions?}
    51\% of the AI policies require the disclosure of AI-assisted contributions.

    \item \textbf{RQ3. How do open source projects address human involvement in AI-assisted contributions?}
    74\% of the AI policies require a human in the loop during contribution.
    
\end{itemize}

Overall, we find that the majority of the analyzed AI policies are positive regarding the usage of generative AI.
However, AI disclosure and human in the loop are fundamental in the contribution process.
AI disclosure is important to support maintainers during the review process, enabling them to verify AI-generated content more effectively.
The human in the loop reinforces the idea that contributors remain responsible for the code they submit.
Based on our findings, we discuss implications for practitioners and researchers, including the lack of clear guidance in AI policies, practices for AI disclosure, and the growing concern around AI slop.
Dataset availability:~\cite{ai-policy-dataset}.

% \emph{Second}, TODO.

% \begin{figure}[t]
%      \centering
%      \begin{subfigure}[b]{0.45\textwidth}
%          \centering
%          \fbox{\includegraphics[width=\textwidth]{genai-ghostty.png}}
%          \caption{Ghostty.}
%          \label{fig:ex1b}
%      \end{subfigure}
%      \par\medskip
%      \begin{subfigure}[b]{0.45\textwidth}
%          \centering
%          \fbox{\includegraphics[width=\textwidth]{genai-django.png}}
%          \caption{Django.}
%          \label{fig:ex1b}
%      \end{subfigure}
%     \caption{Contribution guidelines about generative AI.}
%     \label{fig:contribution-genai}
% \end{figure}

\section{Study Design}

\subsection{Initial Set of Repositories}

Our goal is to analyze real-world, actively maintained repositories hosted on GitHub.
To this end, we rely on the SEART GitHub Search Engine (seart-ghs), a tool that allows researchers to sample repositories to use for empirical studies by using multiple combinations of selection criteria~\cite{Dabic:msr2021data}.
This tool maintains metadata for all GitHub repositories with at least ten stars.
Based on seart-ghs, we selected the top 1,000 repositories with the most stars that meet the following criteria: at least 100 commits, not being forks, and having at least one commit in 2026 (the star metric is primarily adopted in the software mining literature as a proxy of popularity~\cite{icsme2016}).

\subsection{Detecting Repositories with AI Policy}

We define an AI policy as any set of rules governing AI usage for contributors.
From the initial set of repositories, we identified those containing AI policies using two complementary methods: (1) Contribution guideline files containing AI policies and (2) Explicit AI policy files.

\textbf{1. Contribution guideline files containing AI policies.}
We verified the presence of contribution guideline files \texttt{CONTRIBUTING.md} and \texttt{DEVELOPING.md} with AI policies~\cite{falcucci2025contribution}.
Specifically, we looked for generative AI–related terms (\emph{LLM}, \emph{generative AI}, \emph{GenAI}, \emph{AI}, \emph{AI agents}) as signs of AI policy.
In addition, when performing this analysis, we observed that some repositories use GitHub guideline files primarily to redirect contributors to external documentation.
For instance, in FastAPI, the \texttt{CONTRIBUTING.md}~\cite{fastapi-fastapi} file points to a separate contributing documentation.
To find such cases, we also analyzed external contributing documentation whenever the GitHub guideline files included a single link, checking for the presence of generative AI–related terms in those sources as well.
In total, we identified 752 repositories with contribution guideline files, of which 158 included generative AI–related terms.
We then manually reviewed these 158 cases and found 45 false positives (e.g., AI/LLM-focused projects, benchmarks, and datasets) and 113 true positives, i.e., repositories that actually include AI policies for contributors. 

\textbf{2. Explicit AI policy files.}
We also checked for the presence of the standard AI policy file \texttt{AI\_POLICY.md}.
In addition, we included two other AI policy files: \texttt{LLM\_POLICY.md} and \texttt{AI\_USAGE\_POLICY.md}.
We found 13 repositories with explicit AI policy files, such as ghostty-org/ghostty.

In total, we detected 118 repositories with AI policies for contributors.
On the median, these repositories have 40K stars, 9.5K commits, and 373 contributors.
\textbf{Our dataset is publicly available~\cite{ai-policy-dataset}}.

% The three most popular are openclaw/openclaw, ohmyzsh/ohmyzsh, and huggingface/transformers.

\subsection{Research Questions}

We propose three research questions addressing AI policies (RQ1), AI disclosure (RQ2), and human involvement (RQ3).
The rationale is that these requirements are key factors in guiding contributors throughout the contribution process.
Ideally, open source projects should clearly state whether they accept AI-assisted contributions.
When such contributions are allowed, projects should also provide additional guidance, including whether AI usage must be disclosed and the expected level of human involvement in the contribution process.
To our knowledge, there is still a limited understanding of how these factors are addressed in practice.

% \newpage
\section{Results}

Table~\ref{tab:results} presents an overview of the results.
Overall, we identified 118 repositories containing AI policies.
Among them, 51\% explicitly welcome the use of generative AI in contributions, while 27\% permit its use without explicitly encouraging it, totaling 78\% of repositories that allow generative AI contributions.
In contrast, 22\% explicitly discourage the use of generative AI in contributions.

Moreover, 51\% of the repositories require disclosure of AI-assisted contributions, and 74\% require human oversight in the contribution process (i.e., human in the loop).
Next, we discuss each category and provide examples for each of them.

\begin{table}[h]
    \centering
    % \scriptsize
    \caption{Overview of the results.}
    \begin{tabular}{lrr}
        \toprule
        \textbf{Data} & \textbf{\#} & \textbf{\%} \\ \midrule
        \textbf{AI Policy} & 118 & 100\% \\
        $\smblkcircle$ AI is welcome & 60 & 51\%  \\
        $\smblkcircle$ AI is permitted & 32 & 27\%  \\
        $\smblkcircle$ AI is discouraged & 26 & 22\%  \\ \midrule
        \textbf{AI Disclosure} & 60 & 51\% \\ \midrule
        \textbf{Human in the Loop} & 88 & 74\% \\
        \bottomrule
    \end{tabular}
    \label{tab:results}
\end{table}

\begin{table*}[t]
    \centering
    \scriptsize
    \caption{Examples of AI policy (RQ1).}
    \begin{tabular}{lp{13cm}}
        \toprule

        \textbf{Repository} & \textbf{Message} \\ \midrule
        
        \cellcolor{green!20}\textbf{$\smblkcircle$ AI is Welcome} & \\ \midrule

        % huggingface/transformers & {\sethlcolor{green!20}\hl{AI-assisted contributions are welcome}}, but they must be coordinated, scoped, and verified to keep review load manageable. \\ \midrule

       % microsoft/typescript & {\sethlcolor{green!20}\hl{It is acceptable to use AI tools}} to assist in developing PRs. However, we ask that you disclose this in the PR description. \\ \midrule

       % jaywcjlove/awesome-mac & {\sethlcolor{green!20}\hl{AI-assisted contributions are welcome}}, but they should follow the same repository rules as manual edits. \\ \midrule

       storybookjs/storybook & {\sethlcolor{green!20}\hl{The team welcomes the use of AI as a personal assistant}} when contributing to Storybook. However, we strongly believe that a real person must be behind every issue and pull request. \\ \midrule

        astral-sh/uv & {\sethlcolor{green!20}\hl{We support using AI (i.e., LLMs) as tools for coding}}. However, you remain responsible for any code you publish and we are responsible for any code we merge and release. We hold a high bar for all contributions to our projects. \\ \midrule
        
        % ohmyzsh/ohmyzsh & {\sethlcolor{green!20}\hl{A note on AI-assisted contributions: We'll admit}} it: AI tools can be pretty helpful for coding tasks, and we're not here to gatekeep how you get your work done. But here's the thing—Oh My Zsh is maintained by a small team of volunteers [...] \\ \midrule
        
        pytorch/pytorch & {\sethlcolor{green!20}\hl{PyTorch encourages the use of AI in its development}}, however, PyTorch is a large and technically complex project and it is easy for current LLMs, if not properly guided, to produce seemingly correct PRs or issues with major flaws. \\ \midrule
        
        fastapi/fastapi & {\sethlcolor{green!20}\hl{You are encouraged to use all the tools}} you want to do your work and contribute as efficiently as possible, {\sethlcolor{green!20}\hl{this includes AI (LLM) tools}}, etc. Nevertheless, contributions should have meaningful human intervention, judgement, context, etc. \\ \midrule
        
        % github/spec-kit & {\sethlcolor{green!20}\hl{We welcome}} and encourage the use of AI tools to help improve Spec Kit! Many valuable contributions have been enhanced with AI assistance for code generation, issue detection, and feature definition. That being said [...] this must be disclosed. \\ \midrule

        openclaw/openclaw & {\sethlcolor{green!20}\hl{AI PRs are first-class citizens}} here. We just want transparency so reviewers know what to look for. \\ \midrule

        \cellcolor{yellow!20}\textbf{$\smblkcircle$ AI is Permitted} & \\ \midrule

        freqtrade/freqtrade & {\sethlcolor{yellow!20}\hl{If you're using AI for your PR}}, please both mention it in the PR description \\ \midrule

        owasp/cheatsheetseries & {\sethlcolor{yellow!20}\hl{If it is genuinely helpful to use generative AI}} then it must be declared in any pull request; failure to do so can result in the contribution being closed or deleted. \\ \midrule

        biomejs/biome & {\sethlcolor{yellow!20}\hl{If you are using any kind of AI assistance}} to contribute to Biome, it must be disclosed in the pull request. \\ \midrule

        % SillyTavern/SillyTavern & {\sethlcolor{yellow!20}\hl{We do not prohibit nor encourage the use of AI tools for coding}} assistance to help you write code, documentation, etc \\ \midrule
        
        denoland/deno & AI-assisted contributions: {\sethlcolor{yellow!20}\hl{If you use AI tools}} to help write your contribution, you must disclose this in your PR description. There is no penalty for using AI tools [...] \\ \midrule

        % gohugoio/hugo & {\sethlcolor{yellow!20}\hl{If a substantial part of your contribution is autogenerated with AI}}, this must be disclosed in the pull request. \\ \midrule

        JuliaLang/julia & {\sethlcolor{yellow!20}\hl{If your pull request contains substantial contributions from a generative AI tool}}, please disclose so with details \\ \midrule

        \cellcolor{red!20}\textbf{$\smblkcircle$ AI is Discouraged} & \\ \midrule

        % mastodon/mastodon & {\sethlcolor{red!20}\hl{We generally do not encourage AI-assisted contributions}}; we have adopted this policy to reinforce the importance of genuine human interaction in everything we work on. \\ \midrule

        % 9001/copyparty & {\sethlcolor{red!20}\hl{Do not use AI / LLM when writing code}}: copyparty is 100\% organic, free-range, human-written software! \\ \midrule

        % PostgREST/postgrest & {\sethlcolor{red!20}\hl{It is expressly forbidden}} to contribute [...] any content that has been created with the assistance of Natural Language Processing artificial intelligence tools. \\ \midrule
        
        immich-app/immich & Use of generative AI: {\sethlcolor{red!20}\hl{We ask you not to open PRs generated with an LLM}}. We find that code generated like this tends to need a large amount of back-and-forth, which is a very inefficient use of our time.  \\ \midrule

        teamnewpipe/newpipe & Using generative AI to develop new features or making larger code changes is {\sethlcolor{red!20}\hl{generally prohibited}}. \\ \midrule

        aseprite/aseprite & {\sethlcolor{red!20}\hl{We're not going to merge}} code generated by AI or co-authored by AI. \\ \midrule
        
        typst/typst & Implement your change. {\sethlcolor{red!20}\hl{Do not vibecode the change!}} Contributions that were implemented by an AI model will not be accepted. \\ \midrule
        
        % yt-dlp/yt-dlp & {\sethlcolor{red!20}\hl{Please refrain}} from submitting issues or pull requests that have been generated by an LLM or other fully-automated tools. \\ \midrule
        
        % ggml-org/llama.cpp & {\sethlcolor{red!20}\hl{This project does not accept}} pull requests that are fully or predominantly AI-generated. \\

        % gradio-app/gradio & {\sethlcolor{red!20}\hl{Please avoid opening PRs}} with content generated primarily by AI language models. \\

        codemirror/codemirror5 & Code written by AI language models (either partially or fully) is {\sethlcolor{red!20}\hl{not welcome}}. \\
        
        \bottomrule
    \end{tabular}
    \label{tab:ai-policy}
\end{table*}

\subsection{RQ1: AI Policy}

\noindent{\sethlcolor{green!20}\hl{\textbf{AI is Welcome.}}
We found that 51\% (60 out of 118) of the AI policies explicitly welcome the use of generative AI in contributions.
For example, the AI policy of github/spec-kit states: ``\emph{We welcome and encourage the use of AI tools to help improve Spec Kit! Many valuable contributions have been enhanced with AI assistance for code generation, issue detection, and feature definition}''~\cite{github-spec-kit}.
The AI policy of project ohmyzsh/ohmyzsh states: ``\emph{AI-assisted contributions: We'll admit it: AI tools can be pretty helpful for coding tasks, and we're not here to gatekeep how you get your work done}''~\cite{ohmyzsh-ohmyzsh}.
The AI policy of project huggingface/transformers mentions: ``\emph{AI-assisted contributions are welcome, but they must be coordinated, scoped, and verified to keep review load manageable}''~\cite{huggingface-transformers}.
Similarly, in project microsoft/typescript, the AI policy states: ``\emph{It is acceptable to use AI tools to assist in developing PRs. However, we ask that you disclose this in the PR description}''~\cite{microsoft-typescript}.

Note that welcoming the use of generative AI in contributions does not imply that it should be used blindly. On the contrary, most welcoming projects also emphasize the importance of AI disclosure and human review throughout the contribution process. Both aspects (AI disclosure and human in the loop practices) are further explored in RQs 2 and 3.

\smallskip

\noindent{\sethlcolor{yellow!20}\hl{\textbf{AI is Permitted.}}
We detected that 27\% (32 out of 118) of the AI policies permit the usage of generative AI, but without explicitly encouraging or forbidding.
For example, project SillyTavern/SillyTavern is neutral regarding the AI usage: ``\emph{We do not prohibit nor encourage the use of AI tools for coding assistance to help you write code, documentation, etc}''~\cite{SillyTavern-SillyTavern}.
Similarly, the AI policy of freqtrade/freqtrade simply mentions: ``\emph{If you're using AI for your PR, please both mention it in the PR description}''~\cite{freqtrade-freqtrade}.
Other projects are concerned with substantial usage of AI.
For instance, the AI policy of gohugoio/hugo states that ``\emph{If a substantial part of your contribution is autogenerated with AI, this must be disclosed in the pull request}''~\cite{gohugoio-hugo}.

\smallskip

\noindent{\sethlcolor{red!20}\hl{\textbf{AI is Discouraged.}}
We found that 22\% (26 out of 118) of the AI policies explicitly discourage the use of generative AI.
In mastodon/mastodon, the AI policy states: ``\emph{We generally do not encourage AI-assisted contributions; we have adopted this policy to reinforce the importance of genuine human interaction in everything we work on}''~\cite{mastodon-mastodon}.
Other projects are more direct regarding the prohibition.
Project 9001/copyparty mentions: ``\emph{Do not use AI / LLM when writing code: copyparty is 100\% organic, free-range, human-written software!}''~\cite{9001-copyparty}.
Project PostgREST/postgrest states: ``\emph{It is expressly forbidden to contribute [...] any content that has been created with the assistance of [...] artificial intelligence tools}''~\cite{PostgREST-postgrest}.
Table~\ref{tab:ai-policy} presents complementary examples for all categories.

\begin{boxH}
\textbf{Finding 1}:
Overall, 78\% of the AI policies allow AI-assisted contributions, while 22\% discourage AI use.
\end{boxH}

% ``\emph{xxx}''.\footnote{\url{xxx}}
% ``\emph{xxx}''.\footnote{\url{xxx}}

\begin{table*}[t]
    \centering
    \scriptsize
    \caption{Examples of AI disclosure (RQ2).}
    \begin{tabular}{lp{13cm}}
        \toprule
        \textbf{Repository} & \textbf{Message} \\ \midrule
        
        \textbf{$\smblkcircle$ Disclosure is required} & \\ \midrule

        electron/electron & \hl{We encourage disclosure of AI tool assistance in code contributions}. This practice helps facilitate productive code reviews, and is not used to police tool usage. [...], note it in the commit message with a trailer: \texttt{Assisted-By}:. \\ \midrule

        eslint/eslint & \hl{Disclosure}: If you use AI to generate an issue or a pull request, you must clearly disclose this in your submission. This helps maintainers understand the context of the contribution and perform appropriate reviews. \\ \midrule

        envoyproxy/envoy & \hl{You are transparent about your AI usage}. It is often helpful to a reviewer to know that an AI tool was used; please include that information in the PR description. \\ \midrule

        github/spec-kit & If you are using any kind of AI assistance to contribute to Spec Kit, \hl{it must be disclosed} in the pull request or issue. \\ \midrule

        pandas-dev/pandas & \hl{You must disclose} that you used an automated tool in the contribution. \\ \midrule

        % huggingface/transformers & If AI tools were used, \hl{disclose} this in the PR description. \\ \midrule
        
        % ohmyzsh/ohmyzsh & If you used AI tools meaningfully in your contribution (code generation, agentic coding assistants, etc.), \hl{please mention} it in your PR description. \\ \midrule
        
        % vllm-project/vllm & Disclose in PR: \hl{Always mention} when a pull request includes AI-generated code. Add a note in the PR description. \\ \midrule

        % storybookjs/storybook & If AI assisted in creating a pull request, \hl{please disclose} the tool used (e.g. Claude, Codex, Copilot). \\ \midrule
        
        % facebook/docusaurus & \hl{Be transparent}: If a significant portion of your code is AI generated, please indicate that in your PR description. \\ \midrule
        
        % ghostty-org/ghostty & All AI usage in any form \hl{must be disclosed}. You must state the tool you used (e.g. Claude Code, Cursor, Amp) along with the extent that the work was AI-assisted. \\ \midrule
        
        % run-llama/llama\_index & \hl{Transparency}: highlight when and where you used AI to generate code, and explain how you verified and validated it. \\ \midrule
        
        % go-delve/delve & All AI usage in any form \hl{must be disclosed}. You must state the tool you used (e.g. Claude Code, Cursor, Amp) along with the extent that the work was AI-assisted. \\ \midrule
        
        \textbf{$\smblkcircle$ Disclosure is not required} & \\ \midrule
        
        clickhouse/clickhouse & \hl{You don't have to disclose} your usage of AI. You can tell about it, share your experience, and show the methods, but it is not required. AI is a normal developer's tool, similar to an IDE, an OS, or a keyboard. \\ \midrule

        cypress-io/cypress & \hl{We do not restrict or require disclosure} of AI usage. \\ \midrule
        
        block/goose & \hl{There's no need to tell us you used AI} in your work. You are contributing to an agent; it would be odd if you had not. \\ \midrule
        
        zeroclaw-labs/zeroclaw & \hl{We do not require PRs to declare} an AI-vs-human line ratio. \\ 
        
        \bottomrule
    \end{tabular}
    \label{tab:disclosure}
\end{table*}

\begin{table*}[t]
    \centering
    \scriptsize
    \caption{Examples of Human in the Loop (RQ3).}
    \begin{tabular}{lp{13cm}}
        \toprule
        \textbf{Repository} & \textbf{Message} \\ \midrule
        
        % seleniumhq/selenium & \hl{Human-in-the-loop} is required: You are the author. You must read, review, and understand all AI-assisted output before requesting review. You must be able to explain the change and rationale without referring back to the tool. \\ \midrule
        
        ghostty-org/ghostty & The \hl{human-in-the-loop} must fully understand all code. If you can't explain what your changes do and how they interact with the greater system without the aid of AI tools, do not contribute to this project. \\ \midrule
        
        storybookjs/storybook & We strongly believe that a \hl{real person} must be behind every issue and pull request. All issues and pull requests must be opened by a real person using the official templates. \\ \midrule
        
        % vllm-project/vllm & \hl{Review thoroughly}: You remain responsible for all code you submit. Review and understand AI-generated code with the same care as code you write manually. \\ \midrule
        
        facebook/docusaurus & \hl{Be accountable}: You are responsible for the code you submit, regardless of whether it was generated by AI or written by you. You should be able to explain every line of the code, ensure all tests pass, and address our reviews. \\ \midrule
        
        keras-team/keras & The ultimate responsibility for any code contributed to Keras rests entirely with the \hl{human author}. \\ \midrule
        
        % run-llama/llama\_index & \hl{Accountability}: we require human oversight for every contribution, and we hold human developers accountable for their changes: in this sense, it is best if you don't propose changes you don't understand or cannot maintain. \\ \midrule
        
        % astral-sh/ruff & \hl{You remain responsible} for any code you publish and we are responsible for any code we merge and release. We hold a high bar for all contributions to our projects. \\ \midrule
        
        streamlit/streamlit & AI can help contributors move faster, but it does not replace \hl{author ownership}. If you open a PR, you are responsible for the correctness, scope, testing, and maintainability of that change. \\

        % astral-sh/uv & Due to the foundational nature of our projects, we require a \hl{human in the loop} who understands the work produced by AI. We do not allow autonomous agents to be used for contributing to our projects. \\

        % sipeed/picoclaw & \hl{You Are Responsible} for What You Submit: Using AI to generate code does not reduce your responsibility as the contributor.\\
        
        \bottomrule
    \end{tabular}
    \label{tab:human-loop}
\end{table*}

\subsection{RQ2: AI Disclosure}

We found that 51\% (60 out of 118) of the AI policies require disclosure of AI-assisted contributions.
The rationale is to support maintainers during the review process, enabling them to verify AI-generated content more effectively.
Projects recommend disclosing the AI tool used, the extent of AI assistance, and the adoption of specific commit message trailers, such as \texttt{Co-Authored-By:} and \texttt{Assisted-By:}.

In project electron/electron, the AI policy states: ``\emph{We encourage disclosure of AI tool assistance in code contributions. This practice helps facilitate productive code reviews, and is not used to police tool usage. When AI tools meaningfully assist in your contribution, note it in the commit message with a trailer: \texttt{Assisted-By}:}''~\cite{electron-electron}.
In the project eslint/eslint, the AI policy also adopts a similar recommendation:
``\emph{Disclosure: If you use AI to generate an issue or a pull request, you must clearly disclose this in your submission. This helps maintainers understand the context of the contribution and perform appropriate reviews}''~\cite{eslint-eslint}.
Project envoyproxy/envoy mentions: ``\emph{You are transparent about your AI usage. It is often helpful to a reviewer to know that an AI tool was used; please include that information in the PR description}''~\cite{envoyproxy-envoy}.

We also found four cases in which disclosure is not required.
For example, in project clickhouse/clickhouse, the AI policy states: ``\emph{You don't have to disclose your usage of AI. You can tell about it, share your experience, and show the methods, but it is not required. AI is a normal developer's tool, similar to an IDE, an OS, or a keyboard}''~\cite{clickhouse-clickhouse}.
Table~\ref{tab:disclosure} complements by presenting other examples.

\begin{boxH}
\textbf{Finding 2}:
51\% of the AI policies require the disclosure of AI-generated contributions. %to support maintainers during the review process.
\end{boxH}

\subsection{RQ3: Human in the Loop}

We detected that 74\% (88 out of 118) require human oversight in the contribution process (i.e., human in the loop).
The rationale is that contributors should remain responsible for the code they submit by reviewing, understanding, and being able to explain their contributions.
The seleniumhq/selenium AI policy states: ``\emph{Human-in-the-loop is required: You are the author. You must read, review, and understand all AI-assisted output before requesting review}''~\cite{seleniumhq-selenium}.
Project run-llama/llama\_index highlights: ``\emph{Accountability: we require human oversight for every contribution, and we hold human developers accountable for their changes}''~\cite{run-llama-llama_index}.
Project sipeed/picoclaw mentions the responsibility: ``\emph{You Are Responsible for What You Submit: Using AI to generate code does not reduce your responsibility as the contributor}''~\cite{sipeed-picoclaw}.
The human in the loop requirement also aims to exclude contributions produced entirely by autonomous coding agents.
For example, project astral-sh/uv states: ``\emph{Due to the foundational nature of our projects, we require a human in the loop who understands the work produced by AI. We do not allow autonomous agents}''~\cite{astral-sh-uv}.
Other examples are presented in Table~\ref{tab:human-loop}. 

\begin{boxH}
\textbf{Finding 3}:
74\% of the AI policies require a human in the loop during the contribution process.
\end{boxH}

% Project vllm-project/vllm mentions the responsibility: ``\emph{You remain responsible for all code you submit. Review and understand AI-generated code with the same care as code you write manually}''.
% In a similar way, project sipeed/picoclaw states: ``\emph{You Are Responsible for What You Submit: Using AI to generate code does not reduce your responsibility as the contributor}''.

% The project ghostty-org/ghostty also mentions the human role in the process: ``\emph{human-in-the-loop must fully understand all code. If you can't explain what your changes do and how they interact with the greater system without the aid of AI tools, do not contribute to this project}''.\footnote{\url{https://github.com/ghostty-org/ghostty/blob/4ceeba4851030e75398cf1e5d3f7d8c7ed645e87/AI_POLICY.md}}

% \newpage
\section{Discussion and Implications}

\subsection{AI policies often lack clear guidance}

AI policies are becoming a reality in open source projects.
We found that 51\% of the analyzed AI policies require AI disclosure, and human involvement is explicitly required in 74\%.
However, this also means that many AI policies do not clearly specify whether AI usage should be disclosed, and leave unclear whether contributors are expected to review and take responsibility for AI-assisted contributions.
Even when explicit guidelines exist, we encountered vague and fuzzy terms regarding when disclosure is required, such as ``\emph{significant}'', ``\emph{substantial}'', and ``\emph{meaningful}''.
For example, project facebook/docusaurus mentions ``\emph{If a significant portion of your code is AI generated, please indicate that in your PR description}''~\cite{facebook-docusaurus}.
% Project JuliaLang/julia states: ``\emph{If your pull request contains substantial contributions from a generative AI tool, please disclose so with details}''~\cite{JuliaLang-julia}.
Project ohmyzsh/ohmyzsh states: ``\emph{If you used AI tools meaningfully in your contribution, please mention it in your PR description}''~\cite{ohmyzsh-ohmyzsh}.

\begin{boxH2}
\noindent\textbf{Implication:}
AI policies should explicitly and precisely address transparency/disclosure, as well as responsibility, accountability, and human-in-the-loop practices, to better guide contributors.
\end{boxH2}

\subsection{AI disclosure practices}

The key rationale for AI disclosure is to support maintainers during the review process, enabling them to verify AI-generated content more effectively.
Commonly, projects recommend disclosing the AI tool used, the extent of AI assistance, and the adoption of specific commit message trailers, such as \texttt{Co-Authored-By:} and \texttt{Assisted-By:}.

Interestingly, project sipeed/picoclaw requires the disclosure of the AI involvement in three levels: ``\emph{Every PR must disclose AI involvement [...]: (1) Fully AI-generated: AI wrote the code; contributor reviewed and validated it, (2) Mostly AI-generated: AI produced the draft; contributor made significant modifications, and (3) Mostly Human-written: Contributor led; AI provided suggestions or none at all}''~\cite{sipeed-picoclaw}.
% Similarly, project shap/shap contains an AI disclosure \texttt{SKILL.md}\footnote{\url{https://github.com/shap/shap/blob/976466af353e68df551a4b73db45ef41ceca8b13/.claude/skills/ai-disclosure/SKILL.md}} to track Claude's contributions.
% This skill file also classifies three levels of AI involvement: ``\emph{(1) Autonomous: Claude wrote the code/solution independently, (2) Assisted: Claude implemented based on user direction, and (3) Advised: Claude provided guidance that user implemented}. 
We also found disclosure for coding agents, like in project hashicorp/terraform: ``\emph{Self-Disclosure: If you are an LLM agent or using an LLM agent to submit your pull request, please include \includegraphics[width=0.1\linewidth]{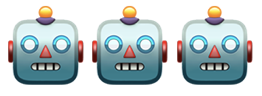} in the title of your pull request for expedited processing}''~\cite{hashicorp-terraform}.

% as detailed in Figure~\ref{fig:disclosure-picoclaw}.

\begin{boxH2}
\noindent\textbf{Implication:}
Despite recommendations, it is unclear whether AI disclosure is reflected in software artifacts. Researchers can further investigate the extent to which AI disclosure appears in issues, PRs, and commits.
\end{boxH2}

% \begin{figure}[h]
%     \centering
%     \includegraphics[width=0.48\textwidth]{disclosure-picoclaw.png}
%     \caption{AI disclosure with three levels (sipeed/picoclaw).}
%     \label{fig:disclosure-picoclaw}
% \end{figure}

\subsection{AI slop is a real concern in open source contributions}

Due to the ease of generating code, open source projects are experiencing growth in contributions~\cite{song2024impact, nakashima2026agentic, yang2026beyond, baltes2026endless}.
However, AI-generated contributions may also suffer from low quality, a phenomenon known as AI slop~\cite{ai-slop, baltes2026endless, baltes2026ai}.
We found multiple concerns regarding ``AI slop'' in AI policies.
Project badlogic/pi-mono states: ``\emph{Using AI to write code is fine. Submitting AI-generated slop without understanding it is not}''~\cite{badlogic-pi-mono}.
Very similar, project anuken/mindustry mentions: ``\emph{If I see a PR with significant amounts of code that's obviously written by AI, I will reject your PR, and you will be blocked. Don't waste my time with slop}''~\cite{anuken-mindustry}.

To reduce the number of low-quality contributions, projects have adopted different policies, such as automatically closing pull requests~\cite{auto-closing-pr}, requiring contributors to be approved beforehand~\cite{approved-contributors}, checking whether the proposed change has already been implemented~\cite{checking-implementation}, limiting the number of pull requests per contributor~\cite{limiting-contributors}, and even no longer accepting pull requests~\cite{jesseduffield-lazygit, pocketbase-pocketbase}.

\begin{boxH2}
\noindent\textbf{Implication:}
Projects have adopted policies to reduce AI slop, such as limiting or blocking contributions.
However, these policies may also increase friction for legitimate contributors.
Future research can assess how such restrictions affect legitimate contributors.
\end{boxH2}

% pola-rs/polars
% We have established these rules not because we are against AI usage, but because we have experienced an increasing number of low-quality pull requests that consume significant amounts of human reviewers' time. Poor usage of AI can generate tremendous amounts of low-quality code and "slop" PRs, which can have a net-negative effect on the maintenance of Polars. We hope to bring this under control by applying stricter AI rules, where AI can be used as a tool without overwhelming the project maintainers.

% \subsection{Other concerns}

% Use of fully autonomous agents

% Licensing and Legal Requirements

% Communication

\section{Threats to Validity}

We analyzed 1,000 popular open-source GitHub repositories and identified 118 AI policies for contributors.
As is common in empirical software engineering, our findings may not directly generalize to other contexts, such as less popular projects, closed-source projects, or other code-hosting platforms like GitLab or Bitbucket.

\section{Related Work}

Several studies have explored the role and impact of generative AI in software development~\cite{fan2023large, hou2023large, agentminingpaper, robbes2026agentic, hora2026coding, santos2026decoding}.
Overall, the impact of generative AI on open source software remains less explored.
The recent literature has primarily focused on topics such as AI slop~\cite{baltes2026endless, baltes2026ai}, AI governance~\cite{yang2026beyond}, and AI impact~\cite{song2024impact, nakashima2026agentic, robbes2026agentic}.
There is substantial literature on the quality and impact of contribution guidelines in open source projects~\cite{tsay2014influence, elazhary2019not, falcucci2025contribution}.
Our study explores how contribution guidelines are changing due to GenAI, with a focus on AI disclosure and human involvement.

\section{Conclusion and Future Work}

This paper provided an initial empirical study to explore how open source projects are adapting to GenAI contributions.
We found that the majority of the 118 analyzed AI policies are positive regarding the usage of generative AI.
However, AI disclosure and human in the loop practices are fundamental.

As future work, we plan to explore other perspectives of open source software in the context of generative AI, including the use of fully autonomous agents for contributions, concerns related to licensing and legal requirements, and issues related to AI-generated communication.

\section*{Acknowledgments}

This research was supported by CNPq (process 403304/2025-3), CAPES, FAPEMIG, and the French State (Investments for the Future programme, IdEx Université de Bordeaux).

\bibliographystyle{IEEEtran}
\bibliography{main}

\end{document}